\documentstyle[aps,multicol,epsf]{revtex}
\begin{document}
\draft
\title{Irreversibility, Mechanical Entanglement and Thermal Melting 
in Superconducting Vortex Crystals with Point Impurities}
\author{Deniz Erta\c s and David R. Nelson}
\address{Department of Physics,
Harvard University,
Cambridge, Massachusetts 02138}
\date{\today}
\maketitle
\begin{abstract}
We discuss the onset of irreversibility and entanglement of vortex lines 
in high $T_c$ superconductors due to point disorder and 
thermal fluctuations using a simplified cage model. 
A combination of Flory arguments, known results from directed polymers
in random media, and a Lindemann criterion are used to estimate the 
field and temperature dependence of irreversibility, 
mechanical entanglement and thermal melting. The qualitative features of 
this dependence, including its nonmonotonicity when disorder is sufficiently
strong, are in good agreement with recent experiments.
\end{abstract}
\pacs{}

\begin{multicols}{2}

\subsection*{Introduction}
The physics of vortex matter in high temperature superconductors (HTSC) has
generated considerable interest since these materials were discovered about 
a decade ago\cite{Bednorz}. The magnetic flux lines in a clean material 
prefer to sit on the sites of a triangular lattice forming an Abrikosov
solid below the superconducting transition temperature $T_c$. 
However, the combined 
effects of thermal fluctuations and various types of disorder, such as oxygen 
vacancies, columnar pins or twin boundaries, significantly alters the phase 
diagram, which is further complicated by typically large 
crystallographic anisotropies\cite{Blatter}. In fact, pinning of flux lines 
by disorder is important technologically, since the flux lines must be 
immobilized in order to eliminate dissipative losses 
associated with flux line motion. In this paper, we will
focus on pinning by pointlike defects, such as oxygen vacancies. 
There are many recent studies in the literature that have considered other 
types of disorder\cite{NelVin,BalKar}.

It is well known that even an arbitrarily small amount of point disorder
is sufficient to disrupt the delicate long range translational order
of the Abrikosov crystal\cite{Larkin}. 
However, the existence of a distinct ``vortex glass'' 
phase\cite{VG,Fisher} at low
temperatures (one which presumably contains dislocations and/or 
disclination defects) is still controversial. Indeed, experiments in which
electron irradiation injects point disorder in twin-free 
YBa$_2$Cu$_3$O$_{7-\delta}$ (YBCO) samples in amounts large enough to 
destroy the first order melting transition have failed to find a sharp
vortex glass transition with universal exponents\cite{Fendrich}.
On the other hand, the presence of a stable, 
topologically ordered, i.e., dislocation-free, ``Bragg glass'' phase 
was recently suggested by various authors\cite{BG} at low magnetic 
fields\cite{footBG} and temperatures in the presence of weak point disorder.
We will loosely refer to such a configuration as the 
``ordered phase''. When dislocations are absent, the interaction 
of a flux line (FL) with its neighbors can be approximated by an effective 
confinement potential that keeps the FL from wandering away,
similar to the Einstein model of phonons in conventional crystals.  
We apply the Lindemann criterion to this 
``cage model'' (where the confinement potential is approximated by a 
simple harmonic well)  in order to investigate the combined
effect of thermal fluctuations and point disorder.
A similar model was used previously to investigate the effects
of correlated disorder due to columnar pins in Ref.~\cite{NelVin}.
The discussion will focus primarily on Bi$_2$Sr$_2$CaCu$_2$O$_{8+\delta}$ 
(BSCCO) crystals, where the large anisotropy confines transitions out of the 
hypothetical Bragg glass to easily accessible fields and temperatures.
We seek to understand the following
three phenomena observed by experiments (See Fig.~\ref{phasediag}): 

\begin{figure}
\narrowtext
\epsfxsize=2.9truein
\vbox{\hskip 0.15truein
\epsffile{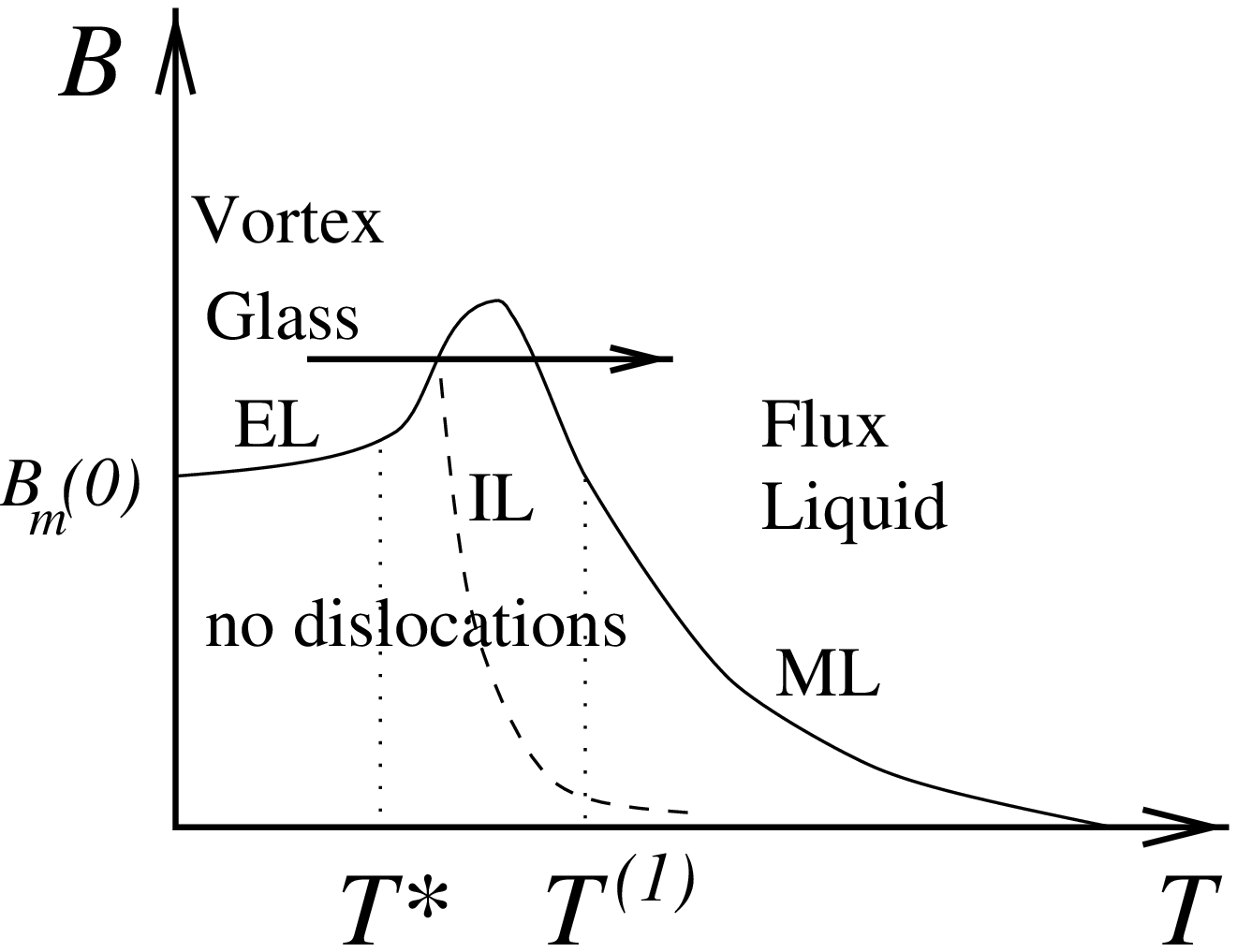}}
\medskip
\caption{Schematic phase diagram indicating the approximate location and
shape of the irreversibility line (dashed) and the combined 
melting--entanglement line (solid). The arrow indicates
the path taken for the situation depicted in Fig.~\protect\ref{fluxline}.}
\label{phasediag}
\end{figure}

{\it (i) The melting line} (ML) marks the loss of order in the 
FL system, and is widely believed to be a first-order thermodynamic
phase transition, most notably marked by a finite jump in the 
magnetization as a function of both temperature and applied
field\cite{meltexp}. 

{\it (ii) The irreversibility line} (IL) marks the cross-over when 
a sample falls out of equilibrium on laboratory time scales. It is 
typically determined by the onset of hysteresis in the magnetization 
curve, and is recently demonstrated to be separate from the melting
line, provided that one excludes effects due to surface barriers\cite{ILexp}. 

\begin{figure}
\narrowtext
\epsfxsize=2.9truein
\vbox{\hskip 0.15truein
\epsffile{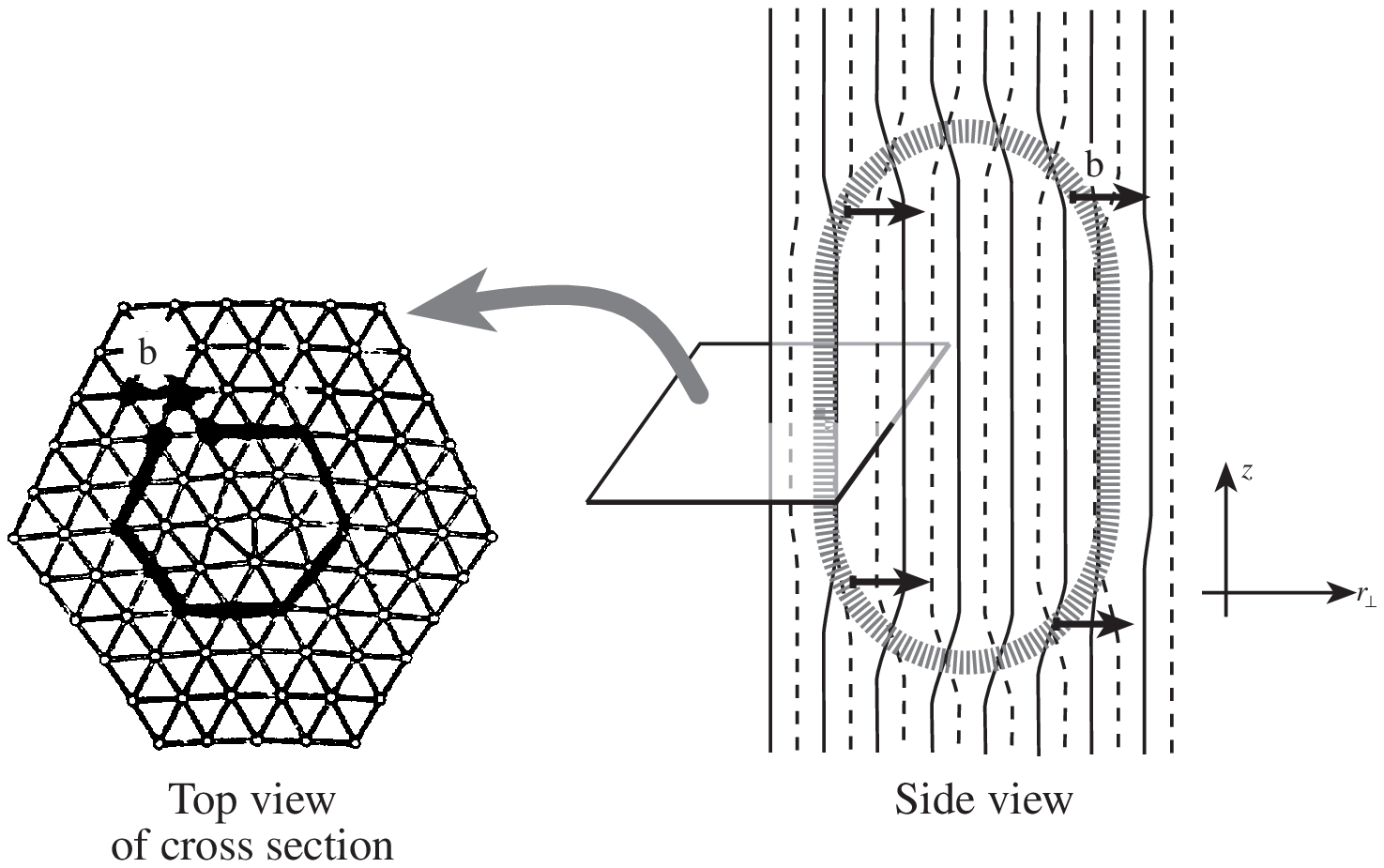}}
\medskip
\caption{Dislocation loop in a topologically ordered vortex crystal.
Nucleation of dislocation pairs in a constant $z$ cross-section
(corresponding to the bottom of the loop) becomes more likely
when the Lindemann criterion is reached.
Dashed lines represent flux lines just behind the plane of the figure.
Top view shows the Burgers' construction applied to a cross-section
through the loop.
dimensions.}
\label{dislocation}
\end{figure}

{\it (iii) The entanglement line} (EL) has only been studied
more recently and can be defined experimentally by a sudden increase 
in the critical current when the field is increased at low 
temperature\cite{Zeldov}. Fairly
independent of temperature, its driving mechanism may be the gradual
proliferation of large dislocation loops favored by point 
disorder\cite{BG,Huse}.
The onset of the large critical currents at low temperatures with increasing
field coincides approximately with the field above which the intensity
of Bragg peaks in neutron diffraction experiments suddenly 
decrease\cite{Cubbit}. Unlike first order melting, these phenomena 
are not accompanied by a magnetization jump, suggesting a continuous
phase transition or a crossover\cite{Zeldov}. The proliferation of
large scale dislocation loops would generate entanglement via their
screw dislocation components (see Fig.~\ref{dislocation}), as suggested
years ago by W\"ordenweber and Kes\cite{Wordenweber} and by
Brandt\cite{Brandt} in the context of dimensional crossover.
Provided line crossing barriers are large, such long wavelength
entanglements would increase the effectiveness of pinning, similar to
ideas about enhanced pinning in arrays of splayed columnar 
defects\cite{Hwa}. As the dislocations become denser with further increases
in the field, vortex crossing angles will become large, flux cutting
easier and critical currents should drop. The critical current in 
experiments does in fact gradually decrease with further increases
in magnetic field\cite{Zeldov}, indicative of a gradual evolution to
a decoupled regime of ``superentangled'' pancake vortices.

Numerical evidence for this mechanical entanglement scenario has been
provided by simulations of Gingras and Huse of a disordered XY model
\cite{Huse}, where vortex rings play the role of dislocation loops
in an Abrikosov lattice. These authors find evidence for a continuous
phase transition to a defect free phase at low temperatures below
a finite disorder strength in three dimensions. In $d=2$, the transition
occurs only at vanishing disorder strength. To make connections with
real vortex arrays, effects of disorder in a given sample are assumed
to increase with increasing magnetic fields. Direct simulations of 
vortex arrays with point disorder in three dimensions by Ryu {\em et al.}
\cite{Ryu} also support this picture, finding evidence for a 
proliferation of lines of disclination pairs (i. e., dislocations)
above a critical filed threshold\cite{footHEX}. 

Our model suggests a relatively simple scenario in which the 
melting, irreversibility and entanglement lines can be understood
and connected in a
way consistent with experimental observations. In this scenario,
a sudden increase in intravalley energy barriers is responsible
for the portion of the IL that lies within the ordered phase.
The lines ML and EL in Fig.~\ref{phasediag} are connected, 
and both signal the proliferation
of dislocations. However, the primary mechanisms are different.
The ML occurs at temperatures above the IL where point disorder 
is ineffective at pinning FLs. 
Melting is thus a competition between thermal fluctuations and
interaction effects. The thermal melting line is only 
slightly perturbed from its position in the absence of disorder.
However, at temperatures below the IL, disorder-induced wandering of 
the FL beyond the Lindemann criterion possibly causes the proliferation 
of {\it frozen-in} dislocations that give rise to a mechanically
entangled configuration with a substantially increased critical
current. This phenomenon, whether a crossover or a sharp continuous
phase transition, results from a competition between interactions
and the effects of point disorder.

For simplicity, let us assume that the average magnetic field $B$ is 
along the c-axis of the sample. The Hamiltonian of a single
FL whose configuration ${\bf r}(z)$ fluctuates
about the line ${\bf r}(z)={\bf r}_0$ can be written 
approximately as  
\begin{equation}
\label{Hamiltonian}
{\cal H}[{\bf r}(z)]=\int_0^L dz \left\{\frac{\tilde\epsilon}{2}
\left(\frac{d{\bf r}}{dz}\right)^2
+V({\bf r}(z),z)+\frac{k}{2}|{\bf r}-{\bf r}_0|^2\right\},
\end{equation}
where $\tilde\epsilon=\epsilon_0/\gamma^2$ is the line tension of 
the FL, $\gamma^2=m_z/m_\perp\gg 1$ is the mass anisotropy,
$\epsilon_0=(\Phi_0/4\pi\lambda)^2$, and $k\approx\epsilon_0/a_0^2$ is
the effective spring constant due to interactions with neighboring
FLs, which are typically at a  distance $a_0=\sqrt{\Phi_0/B}$. 
This approximation to $k$ requires $B > H_{c1}$, so that the vortices
interact with a logarithmic potential at the nearest neighbor spacing.
The probability of
a particular vortex trajectory is proportinal to 
$\exp\{-{\cal H}[{\bf r}(z)]/k_B T\}$. (The Boltzmann constant $k_B$
will be suppressed henceforth.) The disorder
potential $V$ is assumed to have zero mean and short-range correlations
\begin{equation}
\langle V({\bf r},z)V({\bf r'},z')\rangle = \Delta
\delta_\xi^{(2)}({\bf r}-{\bf r'})\delta(z-z'),
\end{equation}
where $\delta_\xi(r)$ is a delta function smeared out to the vortex 
core diameter $\xi$. It will be useful to define a dimensionless disorder
parameter given by 
\begin{equation}
\tilde\Delta=\Delta/( \epsilon_0^2 \xi^3).
\end{equation}

In the {\it absence} of the cage potential $(k=0)$, the model reduces to
a directed polymer in a 2+1 dimensional random medium,
which has been studied extensively\cite{DP}.
At low temperatures, the FL makes transverse excursions
\begin{equation}
u_0^2(\ell)\equiv\langle | {\bf r}(z)-{\bf r}(z+\ell) |^2 \rangle 
\approx \xi^2(\ell/\ell_c)^{2\zeta}, \; \ell > \ell_c,
\end{equation}
and the energy barriers between low-energy configurations grow
as\cite{Drossel}
\begin{equation}
{\cal U}_{p0}(\ell)\approx T^*(\ell/\ell_c)^{2\zeta-1}.
\end{equation}
A Flory-type argument\cite{Nattermann}, combined with the known 
value of $\zeta\approx 5/8$\cite{DP} gives 
\begin{eqnarray}
\ell_c&\approx&(\xi/\gamma)(\tilde\Delta\gamma)^{-1/3}, \\
T^*&\approx&(\xi\epsilon_0/\gamma)(\tilde\Delta\gamma)^{1/3}.
\end{eqnarray} 
At shorter length scales, where it is justified to expand the random 
potential in Eq.(\ref{Hamiltonian}) about ${\bf r}={\bf r}_0$,
\begin{equation}
V({\bf r}(z),z) \approx V({\bf r}_0,z) + ({\bf r}(z)-{\bf r}_0)\cdot
\nabla_{\bf r}V({\bf r}={\bf r}_0,z),
\end{equation} 
the fluctuations are given by the Larkin-Ovchinnikov formula
\begin{equation}
u_0^2(\ell)\approx \xi^2(\ell/\ell_c)^3, \; \ell < \ell_c.
\end{equation}
However, this regime is not relevant to Lindemann discussions of
entanglement provided that $c_La_0>\xi$, and will be ignored henceforth.
($c_L$ is the Lindemann constant, see below.)
Thermal noise is ineffective at depinning the FL from individual
pinning centers as long as the magnitude of thermal fluctuations are 
smaller than quenched fluctuations at the crossover length scale, i.e. 
when $u_{th}^2(\ell_c)=T\ell_c/\tilde\epsilon \ll \xi^2$, or equivalently
when $T \ll T^*$.
At higher temperatures, the excursions of the FL at short distances 
are determined mainly by thermal noise. However, the scaling of
transverse fluctuations is still determined by point disorder
when $\ell> \ell_c(T)$, where the temperature dependent 
crossover length scale is\cite{Blatter}
\begin{equation}
\label{eqlct}
\ell_c(T)=\cases{\ell_c, &$T \ll T^*$,\cr \ell_c\frac{T^*}{T}e^{c(T/T^*)^3}, 
&$T^* \ll T$, } 
\end{equation}
where $c$ is a constant of order unity. The sharp exponential increase
of $\ell_c(T)$ as $T$ exceeds $T^*$ is caused by thermal smearing of
the disorder potential at the marginal dimension $d=3$.  
Transverse fluctuations of the FL are
\begin{equation}
u_0^2(\ell)\approx\cases{\xi^2(\ell/\ell_c)^{5/4}, & $T<T^*,\;
\ell > \ell_c,$ \cr
\frac{T}{\tilde\epsilon}\ell,
& $ T>T^*,\;\ell < \ell_c(T)$, \cr
\frac{T}{\tilde\epsilon}\ell_c(T)\left[\frac{\ell}{\ell_c(T)}\right]^{5/4},
& $ T>T^*,\;\ell > \ell_c(T),$ }
\end{equation} 
and typical energy barriers are
\begin{equation}
{\cal U}_{p0}(\ell)\approx\cases{{T^*}
\left[\frac{\ell}{\ell_c}\right]^{1/4},&  $T < T^*,\; \ell > \ell_c$, \cr
T, & $T > T^*,\; \ell < \ell_c(T)$, \cr
T\left[\frac{\ell}{\ell_c(T)}\right]^{1/4},
&  $T > T^*,\; \ell > \ell_c(T)$.}
\end{equation} 

The main effect of the cage potential is to block
transverse excursions of the FL beyond a confinement length 
$\ell^*\approx\sqrt{\tilde\epsilon/k}=a_0/\gamma$, which is obtained by
balancing the tilting and confinement energies. Every time the vortex
wanders a distance $\ell^*$ along the $z-$axis, it is reflected back 
by the walls of the cage potential and must restart its thermal or
disorder dominated random walk. Thus, the 
mean square displacement of the FL around its equilibrium position
and the typical energy barriers are respectively given by
\begin{eqnarray}
u^2(T)&\approx& u_0^2(\ell^*),  \\
\label{equpt}
{\cal U}_p(T) &\approx& {\cal U}_{p0}(\ell^*).
\end{eqnarray}
The vortex line acts as if it were broken up into independent subsystems
of length $\ell^*$. Similar arguments for the effect of interactions 
on point disorder and thermal fluctuations were given for 
flux {\it liquids} in Ref.~\cite{NelDou}.
  
\subsection*{The Irreversibility Line}

Using the results (\ref{eqlct}-\ref{equpt}), let us investigate
the phenomena described earlier in the context of this simple model.
Naively, one would expect the IL to be associated with a sudden increase
in energy barriers against the motion of individual FLs. When
$T\gg T^{(1)}$, where $T^{(1)}$ is defined by the condition
\begin{equation}
\ell_c(T^{(1)}) = \ell^*,
\end{equation}
pinning centers do not offer a gain in the free
energy of the FLs, and there are no barriers against FL motion. 
The slowest relaxation time of the FL in this regime is
\begin{equation}
\tau_0\approx\frac{\eta}{k}=\frac{8\pi\lambda^2a_0^2}{c^2\xi^2\rho_n},
\end{equation}
where $\eta=\Phi_0 H_{c2}/(\rho_nc^2)$\cite{Tinkham} 
is the friction coefficient of the FL, and $\rho_n$ is the resistivity
of the normal electrons in vortex cores. When $T$ drops into the 
range $T^*<T<T^{(1)}$, 
energy barriers grow rapidly and the relaxation time increases to
\begin{eqnarray}
\tau(T)&=&\tau_0e^{{\cal U}_p(T)/T} \\
&\sim&\tau_0
\exp\left\{\left(\frac{\ell_c(T^{(1)})}{\ell_c(T)}\right)^{1/4}\right\},
\; T^*<T<T^{(1)}.
\end{eqnarray}
Eventually, energy barriers saturate at $T\approx T^*$, resulting in
\begin{equation}
\tau(T)=\tau_0\exp\left\{\frac{T^*}{T}
\left(\frac{\ell^*}{\ell_c}\right)^{1/4}\right\},\; T \ll T^*.
\end{equation}
Thus, in a range of temperatures near $T^*$, the 
relaxation time increases rapidly to macroscopically long times.
The irreversibility condition is satisfied when $\log(\tau(T)/\tau_0)$ 
becomes fairly large. Upon further manipulation, this yields
\begin{equation}
B_{IL}(T)\approx\cases{B_{IL}(T^*)\exp\{-2c(T/T^*)^3\}, &$T>T^*$, \cr
B_{IL}(T^*)(T/T^*)^{-8}, &$T<T^*$.}
\end{equation}

\subsection*{The Melting and Mechanical Entanglement Lines}
In order to estimate where dislocations appear in the $B-T$ phase diagram, 
we can employ a Lindemann criterion as follows:
\begin{equation}
\label{eqLind}
u^2(T_m(B)) = c_L^2 a_0^2,
\end{equation}
where $c_L\approx 0.15 - 0.2$ is the phenomenological Lindemann 
constant\cite{footLin}. 
Thus, for a given sample, the form of $u^2(T)$ will determine the 
shape of the melting line. We next discuss various regimes in this 
function and the corresponding melting fields $B_m(T)$.

1. $T \ll T^*$: In this region,
the Lindemann criterion (\ref{eqLind}) becomes
\begin{equation}
\xi^2(a_0 T^* /\gamma\xi^2\tilde\epsilon)^{5/4} = c^2_L a_0^2.
\end{equation}
Upon simple manipulation, this yields
\begin{equation}
B_m(0) \approx \frac{\Phi_0}{\xi^2}\left(
\frac{\epsilon_0 \xi}{\gamma T^*}\right)^{10/3} c_L^{16/3}, \; T \ll T^*.
\end{equation}
Taking, as parameters for BSCCO, $\xi=20 \AA$, $\gamma=100$, 
$\epsilon_0\xi=1000$ K, and $T^*=10$ K, $c_L\approx0.17$
leads to $B_m(0)\approx 400$ Gauss, in reasonable agreement
with experimental results\cite{Zeldov}. Note that the results 
depend sensitively on the value of the roughness exponent 
$\zeta\approx 5/8$ and the Lindemann constant $c_L$.

2. $T^* < T < T^{(1)}$: In this region, thermal fluctuations
spread the vortex probability distribution and weaken the pinning effect 
of point disorder, but the relevant
length scale at which the cage potential is seen is still determined
by the anomalous wandering characteristics of point pinning. The 
result is
\begin{equation}
B_m(T)\approx B_m(0)\left(\frac{T^*}{T}\right)^{10/3} 
e^{\frac{2c}{3}(T/T^*)^3}, \; T^* < T < T^{(1)},
\end{equation}
indicating an {\it increase} in the melting field with $T$.

3.$T \gg T^{(1)}$: In this region, the FL sees the cage 
potential before it has a chance to take advantage of point disorder,
which is effectively smeared out and can be neglected
as a first approximation. We then find the well-known portion of the 
melting line,
\begin{equation}
B_m(T)\approx\frac{\Phi_0\epsilon_0^2 c_L^4}{\gamma^2T^2}, 
\; T^{(1)} \ll T.
\end{equation}
Note that $\epsilon_0$ has an implicit
dependence on temperature, and the effective anisotropy parameter
$\gamma$ can depend on the amount of point
disorder. The melting field is a decreasing function of temperature
in this regime.

Combining these three regimes gives rise to a nonmonotonic melting field
$B_m(T)$, which is qualitatively depicted in Fig.~\ref{phasediag}. At the
lower temperature side of this peak, $B_m(T)$ lies below the 
irreversibility line, and configurations of the FL are effectively pinned 
by point disorder. Thus, it is more appropriate to 
regard the Lindemann criterion in this regime as a pointer to the onset 
of mechanical entanglement, rather than melting\cite{foot1}. 
More sophisticated methods are needed
to determine the nature (first or second order transition, or crossover) 
of this onset. The high temperature side of the
curve corresponds to the more familiar and studied phenomenon of
flux lattice melting, which was observed as a first-order thermodynamic
phase transition. Finally, the nonmonotonic behavior disappears when
the disorder is too weak, i.e. for $T^* \lesssim 
c_L\xi\epsilon_0/\gamma$.
The transition line observed recently by Khaykovich 
{\it et al.}\cite{Zeldov} in highly anisotropic Bi- based HTSCs is in 
good qualitative agreement with our model.

\subsection*{Numerical Results}
In order to test the analytical results, we have numerically determined
$u^2(T)$ for various values of the magnetic field. 
The general procedure is outlined below, a more detailed discussion 
on Eqs.~(\ref{eqweight}-\ref{eqprob}) can be 
found in Ref.~\cite{NelsonPoly}.

The ``weight''
function for a FL starting from ${\bf r}_i$ at $z=0$ and wandering across
a sample of thickness $\ell$ to  position ${\bf r}$ is given by
\begin{equation}
\label{eqweight}
W({\bf r},{\bf r}_i;\ell)=\int_{{\bf r}'(0)
={\bf r}_i}^{{\bf r}'(\ell)={\bf r}}
{\cal D}{\bf r}'(z) \exp\{-{\cal H}[{\bf r}'(z)]/T\},
\end{equation}
which satisfies a Schr\"odinger-like equation with a time-dependent
(``time $\equiv z$) potential,
\begin{equation}
\label{eqW}
\partial_z W({\bf r},{\bf r}_i;z)=\left[\frac{T}{2\tilde\epsilon}
\nabla_{\bf r}^2-\frac{k}{2}|{\bf r}-{\bf r}_0|^2-V({\bf r}(z),z)
\right]W({\bf r},{\bf r}_i;z).
\end{equation}
For a given realization of point disorder, the weight function
can thus be determined by discretizing and stepping this equation
forward in the time-like parameter $z$.
The probability distribution for the position of a FL at a height 
$z$ which is far from the boundaries of a sample of thickness $\ell$
is proportional to
\begin{equation}
\label{eqprob}
P({\bf r};z) \propto
\int d{\bf r}_i \int d{\bf r}_f W({\bf r}_f,{\bf r};\ell-z)
W({\bf r},{\bf r}_i;z).
\end{equation}

\begin{figure}
\narrowtext
\epsfxsize=2.9truein
\vbox{\hskip 0.15truein
\epsffile{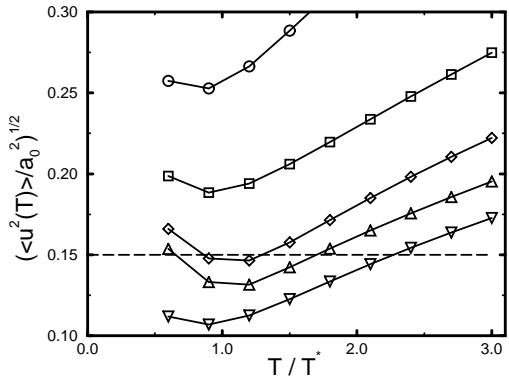}}
\medskip
\caption{Results of a transfer matrix computation of transverse fluctuations
in the cage model. The curves represent temperature scans at
$b\equiv B\xi^2/\Phi_0 =1/100$ (circles), $1/400$ (squares), 
$1/900$ (diamonds), 
$1/1600$ (upwards triangles), and $1/2500$ (downwards triangles). 
The parameters in this run were $\tilde\Delta=\gamma=1$. 
$c_L=0.15$ is shown as a representative Lindemann constant.}
\label{sim}
\end{figure}

For a given realization of disorder, the weight function ratio
$W({\bf r},{\bf r}_i,z)/W({\bf r}',{\bf r}_i;z)$ becomes 
independent of ${\bf r}_i$ when $z\gg\ell^*$ since the FL
forgets its initial position after bouncing off of the confinement
potential a few times.
We have numerically confirmed that the starting position is indeed
indistinguishable for $z > 5\ell^*$, in agreement with the argument 
above. Thus, the integral over initial and final positions in 
Eq.~(\ref{eqprob}) can be eliminated by setting 
${\bf r}_i={\bf r}_f={\bf r}_0$. (Alternatively, one could start from
a uniform weight function.)
Furthermore, the two weight functions that appear in Eq.~(\ref{eqprob})
are statistically independent,
since they traverse different portions of the sample. Taking advantage
of these properties, we have determined $P$ by multiplying two 
weight functions which were propagated
through different samples by $z=5\ell^*$. This probability distribution
was used to determine the fluctuation amplitude, which was subsequently
averaged over different realizations of disorder. 
Fig.~\ref{sim} shows the results for the parameter values 
$\tilde\Delta=\gamma=1$, for different values of the dimensionless 
field $b\equiv B\xi^2/\Phi_0$. In order to solve Eq.~(\ref{eqW}),
each $z-$slice was discretized to a square lattice with lattice
constant $\xi$,
and consecutive $z-$slices were a distance $\delta z=0.05\,\xi/\gamma$
apart. The random potential $V$ was determined independently at each
site of the square lattice and had a correlation length of $\xi/\gamma$ 
along the $z-$direction.

\begin{figure}
\narrowtext
\epsfxsize=2.9truein
\vbox{\hskip 0.15truein
\epsffile{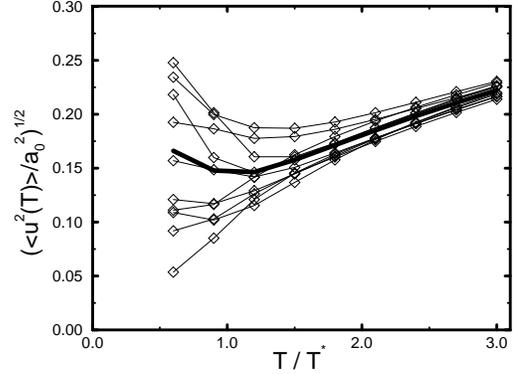}}
\medskip
\caption{Temperature scans of the 10 samples (thin lines) used to 
determine the disorder averaged fluctuation amplitude (thick line)
for $b=1/900$. The spread in data points increases significantly
when $T$ drops below $T^*$.
\label{samples}
}
\end{figure}

\begin{figure}
\narrowtext
\epsfxsize=2.9truein
\vbox{\hskip 0.15truein
\epsffile{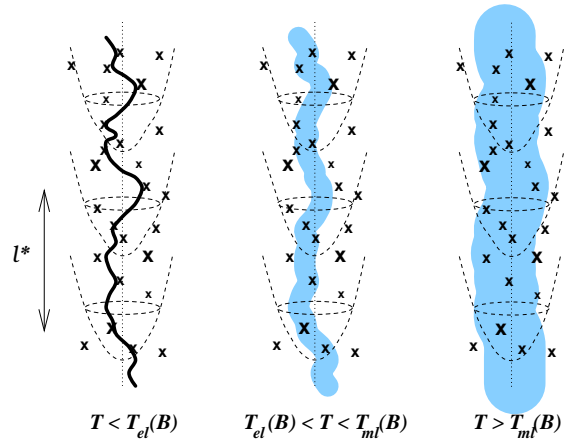}}
\medskip
\caption{Transverse fluctuations of a FL are minimized at an intermediate
temperature (middle) due to the competition between point disorder
and thermal noise. Reentrant behavior is possible if the Lindemann
criterion is crossed during a temperature sweep 
(See Fig.~\protect\ref{sim}). }
\label{fluxline}
\end{figure}

While this method allows exact thermal averaging
of fluctuations, obtaining the disorder average is much more
difficult, as illustrated in Fig.~\ref{samples}. The spread of data
points (each representing one realization of disorder) increases
dramatically for $T<T^*$, when disorder becomes dominant. Ideally,
disorder averaging should be done over many more samples to achieve
reliable results. Nevertheless, reentrant behavior can be deduced from 
the clear nonmonotonicity of the fluctuation amplitude as a function 
of temperature, as seen in Fig.~\ref{sim}. 
This feature can be understood qualitatively
as follows (See Fig.~\ref{fluxline}): As the temperature is increased
from $T=0$, thermal fluctuations will initially act to blur the pinning
centers, thus reducing their pinning strength and giving rise to a 
straighter FL with a smaller fluctuation amplitude. However, further
increases in temperature eventually spread the FL, increasing the
fluctuation amplitude.

\subsection*{Discussion}
In this paper, we have presented the results of a simple ``cage model''
for a FL embedded in a dislocation-free vortex phase. 
The model describes the {\it relative} motion of a vortex with respect
to the cage provided by its neighbors, and is thus insensitive to
the long wavelength disorder fluctuations studied in 
Refs.~\cite{Larkin,VG,Fisher,BG}. An equivalent theory could
presumably be
constructed within the usual Debye phonon model by setting 
$\overline{\left<|{\bf u}({\bf r}+a_0{\bf e}_1)-{\bf u}({\bf r})|^2\right>}
=c_L^2a_0^2$, where ${\bf e}_1$ is a lattice vector and
the brackets and overbar represent thermal and disorder averages
respectively\cite{Fisher}.
We have discussed the sharp increase in energy
barriers (suggestive of an irreversibility line), and the onset of 
instabilities in the ordered phase according to a Lindemann criterion. 
One shortcoming of this model is its inability to verify the 
stability of the ordered phase with respect to large-wavelength 
fluctuations caused by point disorder. However, the stability of the 
Bragg Glass phase with respect to such fluctuations has recently been 
demonstrated\cite{BG}. 
Thus, this model should predict approximately when a given sample 
falls out of equilibrium, and when dislocations proliferate, 
although more sophisticated models are necessary to determine whether
the dislocation proliferation occurs through a phase transition or 
a crossover.

The low temperature entanglement field $B_m(0)$ should not be confused 
with the so called ``decoupling field'' 
$H_\times\approx\Phi_0/(d^2\gamma^2)$ for highly 
anisotropic HTSCs, above which a discretized version of 
Eqs.~(\ref{eqweight}) and (\ref{eqW}) should be used\cite{decouple}.
vortex line behave roughly independently in each Cu-O layer. 
The ratio of these two fields is
\begin{equation}
\frac{B_m(0)}{H_\times}\approx  
\gamma^2c_L^{16/3}\left(\frac{d}{\xi}\right)^2
\left(\frac{\epsilon_0\xi}{\gamma T^*}\right)^{10/3}.
\end{equation}
For materials such as YBCO or BSCCO, this ratio is typically less
than 1. A more detailed account of layering effects and the dispersive 
nature of the line tension is beyond the scope of this work.

We are indebted to E.~Zeldov for stimulating our interest in this problem.
We also benefited from helpful conversations with D.~Huse, P.~Le~Doussal,
C.~Lieber and V.~Vinokur. 
This work was supported by the National Science Foundation, primarily
by the MRSEC program through Grant DMR-9400396 and in part through
Grant DMR-9417047.

\end{multicols}


\begin{references}
\bibitem{footBG}By ``low'' magnetic fields, we mean fields $B$ smaller
than the ``decoupling field'' $H_\times \approx \Phi_0/(d^2\gamma^2)$,
where $\gamma^2=m_z/m_\perp \gg 1$ is the mass anisotropy, 
$\Phi_0 = hc/2e$ is the flux quantum, and $d$ is the spacing between 
copper oxide planes. The fields considered here are nevertheless assumed
to be large enough so that $a_0 \le \lambda$, where $a_0=\sqrt{\Phi_0/B}$
is a typical vortex spacing and $\lambda$ is the London penetration
depth.
\bibitem{footLin}It should be stressed that, although useful for locating
important physical phenomena in the phase diagram, the Lindemann criterion
by itself cannot predict whether a given transition is first order, second
order, or simply a crossover.
\bibitem{foot1}Since the mechanisms that generate dislocations are different,
the Lindemann parameters may not be identical for the melting and entanglement
lines.
\end{references}
\end{document}